\documentclass[a4paper, oneside, twocolumn, notitlepage, 10pt]{extarticle_ecoc}
\usepackage{ecoc}
\usepackage{acronym}
\usepackage{orcidlink}

\acrodef{ReAct}{reasoning and act}
\acrodef{LLM}[LLM]{large language model}
\acrodef{DT}[DT]{digital twin}
\acrodef{TAPI}[T-API]{Transport Application Programming Interface}
\acrodef{MCP}[MCP]{Model Context Protocol}
\acrodef{OPM}[OPM]{optical performance monitoring}
\acrodef{QoT}[QoT]{quality-of-transmission}
\acrodef{ROADM}[ROADM]{reconfigurable optical add-drop multiplexer}
\acrodef{SIP}[SIP]{service-interface-point}
\acrodef{UUID}[UUID]{universally unique identifier}
\acrodef{HTTP}[HTTP]{Hypertext Transfer Protocol}
\acrodef{SDN}[SDN]{software-defined networking}
\acrodef{ZSM}[ZSM]{Zero-touch network and Service Management}

\begin{document}
\selectlanguage{english}


\title{A T-API-Compliant ReAct Agentic Loop for Optical Networks: Generic vs. Domain-Specific Tool Abstractions}%


\author{
    Morteza Ahmadian~\orcidlink{0000-0002-1251-6288}, Paolo Monti~\orcidlink{0000-0002-5636-9910},
    Carlos Natalino~\orcidlink{0000-0001-7501-5547}
}

\maketitle


\begin{strip}
    \begin{author_descr}

        Department of Electrical Engineering, Chalmers University of Technology, Gothenburg, Sweden.
        \textcolor{blue}{\uline{seyedah@chalmers.se}}, \textcolor{blue}{\uline{mpaolo@chalmers.se}}, \textcolor{blue}{\uline{carlos.natalino@chalmers.se}}

    \end{author_descr}
\end{strip}


\begin{strip}
    \begin{ecoc_abstract}
        Optical networks need intent-driven, closed-loop agentic management, a key enabler for higher autonomy levels. We present the first T-API-compliant reasoning and act (ReAct) loop. We show that domain-specific composite tools achieve 90\% oracle-validated correctness with threefold token savings compared to generic tools.~\textcopyright~2026 The Authors
    \end{ecoc_abstract}
\end{strip}


\section{Introduction}

The telecommunications industry is still operating at low autonomy levels (L2--L3), while higher levels (e.g., L4) are targeted for 2030~\cite{Webb_2024_AutonomousNetworksSearch}.
In this context, both IMT-2030 and the ETSI \ac{ZSM} framework~\cite{ITUR_2023_FrameworkOverallObjectives, __GSZSM016} establish intent-driven, closed-loop, agentic operations as the target operating model.
Meanwhile, \acp{DT} have emerged as safe sandboxes in which actions proposed by \acp{LLM} can be validated before being instantiated in production networks~\cite{Wang_2021_RoleDigitalTwin}.

The \ac{TAPI} is a standard, vendor-neutral northbound interface for multi-vendor, disaggregated optical network control.
Nevertheless, agentic-LLM work in optical networking has so far targeted device-level YANG models~\cite{Liu_2025_FirstFieldTrial, Liu_2026_FieldTrialLLMpowered, Zaid_2025_MultiAgentDesignLLMassisted, Shariati_2025_DataSovereignLLMAssisted, Zhou_2024_LargeLanguageModelDriven}, vendor specific \ac{SDN} APIs~\cite{Huang_2025_FieldTrialLLMBased, Xu_2025_CrossDomainOrchestrationMultiAgent, Wang_2026_AgenticAIScalable}, and proprietary simulators~\cite{Jiang_2024_OptiCommGPTGPTbasedVersatile, Th2A.33,M4A.6}, leaving the standardized, vendor-neutral T-API northbound unexplored.
LLM-driven optical control has been demonstrated in field trials over NETCONF/YANG with plan-and-execute or \ac{ReAct} sub-loops~\cite{Liu_2026_FieldTrialLLMpowered, Huang_2025_FieldTrialLLMBased}, over OpenConfig/NETCONF with fixed multi-agent pipelines~\cite{Zaid_2025_MultiAgentDesignLLMassisted, Shariati_2025_DataSovereignLLMAssisted}, and over vendor-specific SDN APIs with multi-agent orchestration~\cite{Xu_2025_CrossDomainOrchestrationMultiAgent}. Simulator-only and tutorial works complete the picture~\cite{Jiang_2024_OptiCommGPTGPTbasedVersatile, Th2A.33}. Only~\cite{Jafari_2025_LLMAssistantTAPI} has explicitly addressed T-API, and did so as a vendor-code mediator with a fixed classifier-generator-validator pipeline, not as a ReAct agent for autonomous management.

Graph-based agents provide a deterministic set of steps that the agent had to go through, including the execution of heuristics. Meanwhile, ReAct-based agents do not follow a pre-defined set of steps, and can decide which set of steps to take, tools to call, or to interact with the user by e.g. asking for further details.

In the meantime, tool abstraction has been studied contrasting code versus structured outputs~\cite{Wang_2024_NetConfEvalCanLLMs}, \ac{MCP} versus code generation~\cite{Wang_2026_AgenticAIScalable}, and generic application interfaces~\cite{Qin_2024_ToolLLMFacilitatingLarge}.
First attempts at standardized benchmarks exist~\cite{Zhang_2026_AutoONBenchBenchmarkLarge, Wang_2024_WhenLargeLanguage}, but they rely on LLM-as-judge or human rubrics, and hallucination is at best qualitatively catalogued~\cite{Jiang_2024_OptiCommGPTGPTbasedVersatile, Yuan_2026_EnhancingLargeLanguage}.
\ac{MCP} for network management has been proposed but neither implemented nor measured~\cite{YUANYUANYANG_2026_ApplicabilityMCPNetwork}.
Once the LLM is capable enough (14B-class and above according to current studies~\cite{Th2A.33, Wang_2026_AgenticAIScalable}), the dominant lever on success rate and efficiency shifts from model capability to the tool layer.
Nevertheless, no prior work measures how the tool abstraction itself affects correctness, token efficiency, and hallucination on the same agent.

To this end, we present, to the best of our knowledge, the first T-API-compliant ReAct agentic loop for optical networks, together with the first controlled comparison of generic \ac{HTTP} tools against a domain-specific T-API tool layer (12 atomic and 4 composite operations) on a common agent and scenario suite.
We evaluate 10 scenarios covering querying, analysis, provisioning, and multi-turn workflows, with an automated oracle that separates execution success from answer correctness and identifies each agent failure.
Results demonstrate that domain-specific tools significantly increase correctness, reduce token usage, and suppress hallucinations relative to raw protocol interaction, with the abstraction effect exceeding the model-choice effect. The proposed framework constitute a step toward principled studies of future agentic T-API adapters.


\section{T-API-Compliant ReAct Agentic Architecture}

\begin{figure}[t]
    \centering
    \includegraphics[width=\linewidth]{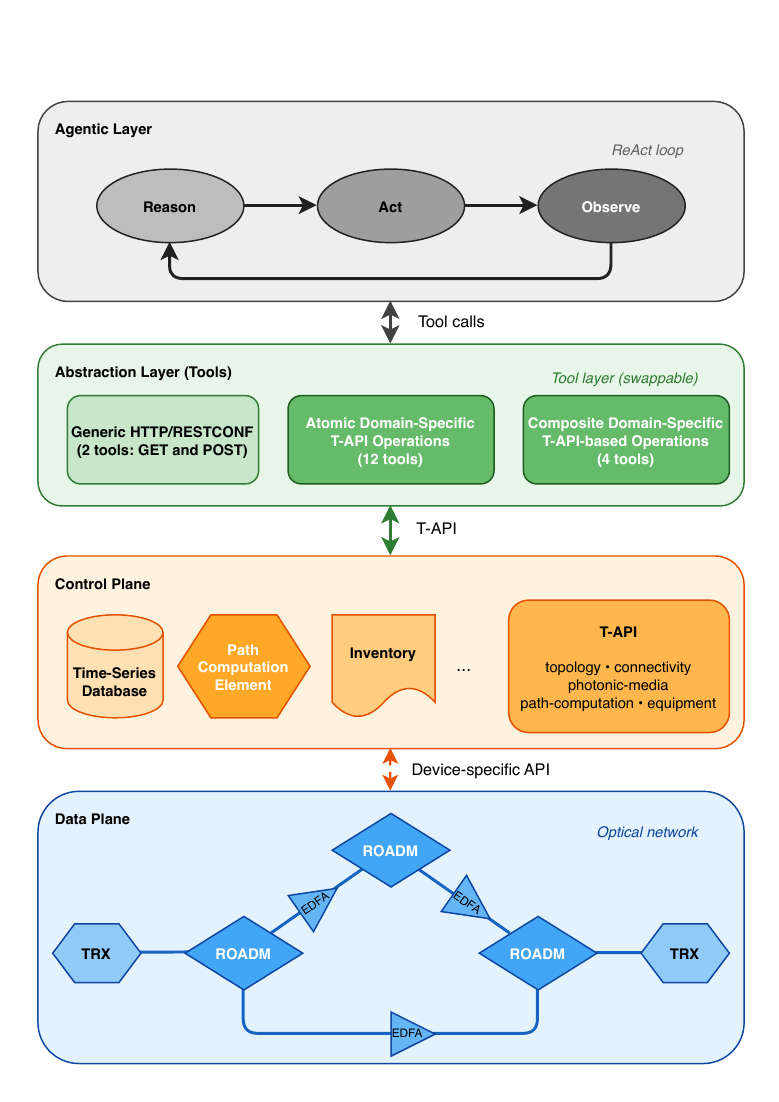}
    \caption{The T-API-Compliant ReAct Agentic Architecture.}
    \label{fig_architecture}
\end{figure}

Fig.~\ref{fig_architecture} illustrates the proposed architecture.
A ReAct reasoning loop drives a tool layer that targets a T-API RESTCONF northbound, covering topology, connectivity, photonic-media, path-computation, and equipment contexts.
The tool layer can be chosen between generic and domain-specific abstractions without altering either the agent or the northbound, which isolates the abstraction as the sole independent variable.

The ReAct agent layer is implemented on top of an LLM runtime.
Unlike fixed multi-agent pipelines \cite{Jafari_2025_LLMAssistantTAPI, Zaid_2025_MultiAgentDesignLLMassisted, Shariati_2025_DataSovereignLLMAssisted} and plan-and-execute prior work \cite{Liu_2026_FieldTrialLLMpowered, Wang_2024_WhenLargeLanguage}, our loop iteratively reasons, calls a tool, observes, and decides dynamically, according to instructions.

Two tool alternatives share the same agent and the same control and data planes.
The first alternative employs generic HTTP tools, that is, two primitives (GET/POST with URL and body) following the philosophy of~\cite{Qin_2024_ToolLLMFacilitatingLarge}, where the LLM must construct and interpret T-API paths and messages by itself.
The second alternative exposes 16 typed operations (12 atomic and 4 composite) covering topology browsing, path computation, modulation selection, connectivity lifecycle, and GNPy-based \ac{QoT} estimation and \ac{OPM} telemetry.


\section{Experimental Setup}

Our setup uses a \ac{DT} acting as the northbound T-API.
The \ac{DT} combines a GNPy-based QoT engine~\cite{Curri_2022_GNPyModelPhysical}, a NetworkX topology and inventory, and routing based on~\cite{Natalino_2024_OpticalNetworkingGym}.
Our results are obtained using the CORONET CONUS topology (75 \acp{ROADM} and 198 fiber spans) operated in the C-band, according to the GNPy-provided example configuration files.

Three tool abstractions targeting the same northbound are evaluated, varying the level of T-API specific manipulation required by the LLM powering the agent.
The generic HTTP/RESTCONF option exposes two primitives: \texttt{http\_request} (read-only GET) and \texttt{http\_request\_mutate} (POST/PUT/PATCH/DELETE).
Under this option the LLM constructs RESTCONF paths and JSON, resolves \ac{SIP} \acp{UUID}, navigates raw YANG-model JSON, and composes POST bodies itself.

The single-call tools abstraction comprises 12 atomic operations, each of which maps 1:1 to a T-API endpoint or performs pure in-process filtering over another atomic tool's output, namely \texttt{list\_nodes}, \texttt{get\_node\_details}, \texttt{get\_link\_details}, \texttt{list\_services}, \texttt{get\_service\_details}, \texttt{get\_service\_path}, \texttt{get\_equipment}, \texttt{get\_spectrum\_overview}, \texttt{get\_opm\_service}, \texttt{find\_degraded\_services}, \texttt{estimate\_qot}, and \texttt{delete\_service}.

The multi-call tools abstraction adds 4 composite operations on top of the single-call set: \texttt{get\_topology\_summary}, \texttt{compute\_paths}, \texttt{find\_best\_modulation} (cascaded DP-64QAM to DP-16QAM to DP-QPSK via up to three \texttt{estimate\_qot} calls), and \texttt{provision\_service}.
These tools make it easier for the agent to execute operations that naturally require multiple interactions with the network through the T-API.

The scenario suite contains 10 scenarios spanning querying (Q1--Q4), analysis (A1--A3), provisioning (P1--P2), and multi-turn flows (MT1). It spans the lifecycle categories of \cite{Zhang_2026_AutoONBenchBenchmarkLarge} and \cite{Liu_2026_FieldTrialLLMpowered} while being grounded in T-API semantics.
Node-dependent scenarios randomize source and destination pairs on every run by sampling live SIP endpoints from the topology. Representative cases include A1 (best modulation query on a routable path), A2 (modulation feasibility on potentially long, high-loss paths), A3 (k-shortest paths with comparative QoT estimates), P1 (lightpath provisioning after a QoT feasibility check), P2 (400G service attempt on a potentially infeasible path), and MT1 (explore-then-act across two turns).

Consistent with the data-sovereign motivation of \cite{Shariati_2025_DataSovereignLLMAssisted, Th2A.33, Huang_2025_FieldTrialLLMBased}, four open on-premises \acp{LLM} from the Qwen family are evaluated: Qwen2.5-32B-Instruct (dense transformer fine-tuned for tool calling, primary baseline at temperature 0.0), Qwen3.5-35B (mixture-of-experts model activating only a sparse subset of parameters per token), Qwen3.5-9B (dense mid-range model), and Qwen3.5-4B (compact dense model stress-testing resource-constrained deployments).
Each LLM, tool abstraction, and scenario combination is executed for 20 runs for statistical significance.
The agent is implemented with LangGraph, and all inference runs locally via vLLM on a single NVIDIA RTX 6000 GPU with 96~GB.

Two metrics are collected per run.
The success rate reports oracle-validated answer correctness, combining tool-presence checks with value checks against ground-truth quantities fetched from the \ac{DT}.
The oracle further types each failure as zero-tool, wrong-value, wrong-modulation, or missing-grounding, thereby extending the qualitative taxonomies of~\cite{Jiang_2024_OptiCommGPTGPTbasedVersatile} and~\cite{Yuan_2026_EnhancingLargeLanguage}.
Provisioning is scored in two stages: a presence layer confirming that \texttt{provision\_service} returned a service UUID (HTTP 201), and an oracle-correctness layer verifying modulation selection against the pre-computed \texttt{best\_modulation} and, on infeasible paths, the agent's ability to refuse provisioning.
Token usage captures the total number of tokens sent to and produced by the \ac{LLM} as a measure of cost of the operations.
Note that in our case, as well as in real-world scenarios, token usage is accounted for regardless of agent success.


\section{Numerical Results}

\begin{figure*}[t]
    \centering
    \includegraphics[width=\linewidth]{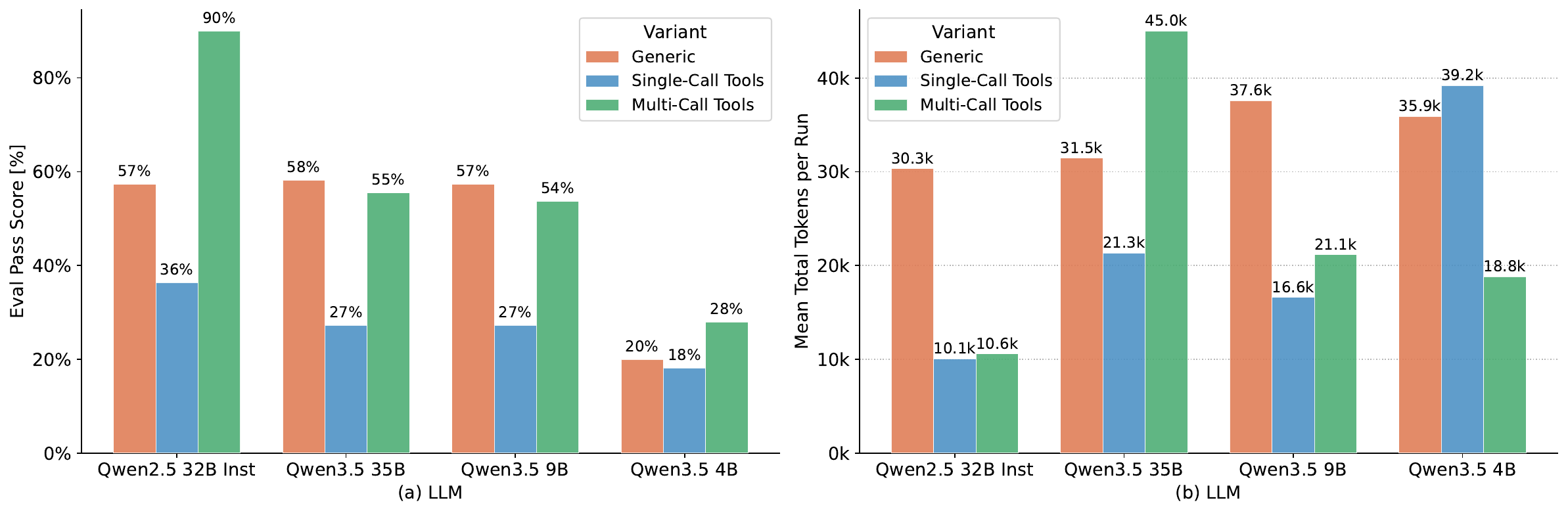}
    \caption{Evaluation pass rate (a) and mean total tokens per run (b) for different LLM models and tool abstractions.}
    \label{fig_results}
\end{figure*}

Fig.~\ref{fig_results}(a) shows the success rate and Fig.~\ref{fig_results}(b) the mean total number of tokens per run across the four LLMs and three tool abstractions.
The results reveal that, under the tested conditions, both tool abstraction and \ac{LLM} are decisive on both correctness and cost.

The clearest signal comes from the generic tool abstraction, where the pass rate flattens at approximately 57--58\% across the larger models, despite nearly a fourfold spread in \ac{LLM} size.
Token usage is uniformly high (30k--38k per run).
The success rate of the smallest model collapses to 20\%.
Therefore, the solution using generic tool abstraction uses more tokens without higher success, and saturates success rate below 60\% regardless of scale.

Against this saturation plateau, the enhanced multi-call abstraction unlocks a decisive improvement: Qwen2.5-32B-Instruct improves from 57\% under generic tools to approximately 90\% under composite tools -- a gain exceeding 30\% -- while its token usage drops to roughly 10.6k per run, among the lowest in the whole experiment. Tool-calling fine-tuning pays off simultaneously on both axes.
This result confirms that the size-vs-grounding observations of~\cite{Th2A.33} and~\cite{Wang_2024_WhenLargeLanguage} are also observed here.
The results suggest that for T-API agentic control, combining an instruction-tuned tool-calling LLM with typed composite tools dominates over \ac{LLM} size.

The other models, however, show that composite tools alone are not a sufficient for \acp{LLM} not prepared for tool calling.
Exposing atomic single-call tools still forces the agent to orchestrate multi-step T-API flows (SIP resolution, \ac{QoT} estimation, cascaded modulation screening, path cross-comparison), and non-instruct mid-range models (Qwen3.5-9B, Qwen3.5-35B) collapse under this orchestration burden.
The success rate roughly halves relative to the instruction-tuned model on composite tools, and Qwen3.5-35B consumes the highest number of tokens among the tested setups, with up to 45k tokens.
Qwen3.5-4B remains below 28\% success rate on every abstraction regardless of token usage, indicating a floor on capability below which tool-layer design cannot compensate.
Composite tools absorb recurring flows into deterministic logic, but only models fine-tuned to invoke them reliably translate that abstraction into correctness gains.

\section{Conclusions}

We introduced the first T-API-compliant ReAct agentic loop for optical networks.
We benchmarked generic vs. domain-specific tool abstractions through an oracle validator with a domain-specific procedures.
The results showed that agents become operationally viable once grounded in T-API-aware tools, but need to be based on \acp{LLM} optimized for tool calling.
The proposed architecture combining a ReAct agent and domain-specific tools whose signatures map directly onto T-API objects, can be dropped into any T-API-compliant domain controller.
Future work may target a T-API MCP server that mimics our 16 domain-specific tools but allowing remote execution.
Porting the tool layer to MCP would test whether the MCP wire format would preserve the correctness gains measured in this work.

\newpage
\section{Acknowledgements}
The work presented in the paper originated from the EUREKA CELTIX-NEXT SUSTAINET-Advance Flagship project supported (in Sweden) by the Sweden’s Innovation Agency VINNOVA (Dnr: 2025-02987).

\printbibliography

\vspace{-4mm}

\end{document}